\documentclass[12pt,english,manuscript,aps,prd,A4,English, 12pt]{revtex4-1}
\usepackage[T1]{fontenc}
\usepackage[latin9]{inputenc}
\usepackage{geometry}
\usepackage{color}
\usepackage{babel}
\usepackage{amstext}
\usepackage{esint}
\usepackage[unicode=true,pdfusetitle,
 bookmarks=true,bookmarksnumbered=false,bookmarksopen=false,
 breaklinks=false,pdfborder={0 0 0},pdfborderstyle={},backref=false,colorlinks=true]
 {hyperref}
\hypersetup{
 linkcolor=blue, citecolor=cyan}

\makeatletter

\usepackage{xcolor}

\makeatother

\begin{document}
\title{Analytic solutions of relativistic dissipative spin hydrodynamics
with Bjorken expansion}
\author{Dong-Lin Wang}
\email{wdl252420@mail.ustc.edu.cn}

\affiliation{Department of Modern Physics, University of Science and Technology
of China, Anhui 230026, China}
\author{Shuo Fang}
\email{fangshuo@mail.ustc.edu.cn}

\affiliation{Department of Modern Physics, University of Science and Technology
of China, Anhui 230026, China}
\author{Shi Pu}
\email{shipu@ustc.edu.cn}

\affiliation{Department of Modern Physics, University of Science and Technology
of China, Anhui 230026, China}
\begin{abstract}
We have studied analytically the longitudinally boost-invariant motion
of a relativistic dissipative fluid with spin. We have derived the
analytic solutions of spin density and spin chemical potential as
a function of proper time $\tau$ in the presence of viscous tensor
and the second order relaxation time corrections for spin. Interestingly,
analogous to the ordinary particle number density and chemical potential,
we find that the spin density and spin chemical potential decay as
$\sim\tau^{-1}$ and $\sim\tau^{-1/3}$, respectively. It implies
that the initial spin density may not survive at the freezeout hyper-surface.
These solutions can serve both to gain insight on the dynamics of
spin polarization in relativistic heavy-ion collisions and as testbeds
for further numerical codes.
\end{abstract}
\maketitle

\section{Introduction \label{sec:Introduction}}


In the non-central relativistic heavy ion collisions, huge orbital
angular momenta are generated and polarize the particles in the quark
gluon plasma (QGP) through the spin-orbital couplings \citep{ZTL_XNW_2005PRL,ZTL_XNW_2005PLB}.
In 2017, the global polarization of $\Lambda$ and $\bar{\Lambda}$
hyperons led by the initial huge orbital angular momentum has been
measured by STAR experiments and been well understood by various phenomenological
models \citep{Karpenko_epjc2017_kb,XieYilong_prc2017_xwc,LiHui_prc2017_lpwx,Sun:2017xhx,WeiDexian_prc2019_wdh,Shi_plb2019_sll,Fu:2020oxj,Ryu:2021lnx}.
The experimental results also indicate that QGP is the most vortical
fluid \citep{STAR:2017ckg} so far.


The experimental data in Au+Au collisions at $200$ GeV for the local
spin polarization, which is the azimuthal angle dependent spin polarization
of the $\Lambda$ and $\bar{\Lambda}$ hyperons along the beam and
the out-plane directions \citep{Adam:2019srw}, disagrees with many
phenomenological models, e.g. relativistic hydrodynamics model \citep{Becattini_prl2018_bk,Fu:2020oxj}
and transport models \citep{XieYilong_prc2017_xwc,Xia:2018tes,WeiDexian_prc2019_wdh}.
Later on, people find that this disagreement cannot be solved by considering
the feed-down effects \citep{Xia:2019fjf,Becattini_epjc2019_bcs}.
Although the kinetic theory of massless fermions in Ref. \citep{Liu:2019krs}
and results from a simple phenomenological model in Ref. \citep{Voloshin_2018epjWeb}
can show the similar azimuthal angle dependence as experimental data,
it is still a puzzle in the relativistic heavy ion community.

To solve this puzzle, it requires a deeper understanding of spin effects
in the QGP from both microscopic and macroscopic theories. One microscopic
theory for massive fermions is the quantum kinetic theory (QKT) for
massive fermions, which is a natural extension of the chiral kinetic
theory for massless fermions \citep{Stephanov:2012ki,Son:2012zy,Chen:2012ca,Manuel:2013zaa,Manuel:2014dza,Chen:2014cla,Chen:2015gta,Hidaka:2016yjf,Mueller:2017lzw,Hidaka:2018mel,Huang:2018wdl,Gao:2018wmr,Liu:2018xip,Lin:2019ytz,Lin:2019fqo}.
Recently, the QKT for massive fermions with the collisional kernel
has been developed \citep{Gao:2019znl,Weickgenannt:2019dks,Weickgenannt:2020aaf,Weickgenannt:2020sit,Hattori:2019ahi,Yang:2020hri,Liu:2020flb,Weickgenannt:2021cuo,Sheng:2021kfc}.
Also see the early works on the statistical model for the relativistic
particles with spin \citep{Becattini:2007nd,Becattini:2007sr,Becattini:2013fla}
and Ref. \citep{Zhang:2019xya} for a microscopic model for spin polarization
through particle collisions.

Very recently, many studies have shown that the polarization can also
been induced by shear viscous tensor, e.g. from early studies of massless
fermions \citep{Hidaka:2017auj}, the recent studies for massive fermions
\citep{Liu:2020dxg,Liu:2021uhn} and statistic model \citep{Becattini:2021iol}.
Those studies \citep{Fu:2021pok,Becattini:2021iol,Yi:2021ryh} show
quantitative agreement with experimental data by adding the shear
induced polarization to the original studies \citep{Becattini:2013fla,Fang:2016uds}.


The macroscopic theory for the spinness particles is the relativistic
spin hydrodynamics, which is a theory with the relativistic hydrodynamics
equations coupled to the conservation equation of total angular momentum.
The expression of spin hydrodynamics has been derived from the entropy
principle \citep{Hattori:2019lfp,Fukushima:2020qta,Fukushima:2020ucl,Li:2020eon,She:2021lhe},
the Lagrangian effective theory \citep{Montenegro:2017lvf,Montenegro:2017rbu},
kinetic approaches \citep{Florkowski_prc2018_ffjs,Florkowski:2018myy,Becattini_plb2019_bfs,Florkowski:2018fap,Bhadury:2020puc,Shi:2020qrx},
and general discussion from field theory \citep{Gallegos:2021bzp}.
The ideal spin hydrodynamics has been discussed in Ref. \citep{Florkowski_prc2018_ffjs,Florkowski:2018myy,Becattini_plb2019_bfs,Florkowski:2018fap,Hattori:2019lfp,Bhadury:2020puc,Shi:2020qrx,Montenegro:2017lvf,Montenegro:2017rbu,Li:2020eon}.
In the gradient expansion, the dissipative spin hydrodynamics are
derived in canonical \citep{Hattori:2019lfp} and Belinfante form
\citep{Fukushima:2020qta,Fukushima:2020ucl}. Also, see recent reviews
\citep{Wang:2017jpl,Becattini:2020ngo,Becattini:2020sww,Gao:2020vbh,Liu:2020ymh}
and the reference therein.


Analogous to works on the relativistic magnetohydrodynamics \citep{Pu:2016ayh,Roy:2015kma,Pu:2016bxy,Pu:2016rdq,Siddique:2019gqh,Wang:2020qpx},
it is necessary to derive the analytic solutions for the dissipative
spin dynamics. These analytic solutions provide the power law behavior
of the quantities related to the spin as a function of proper time
$\tau$, such as the spin density and spin chemical potential. On
the other hand, the numerical codes for the dissipative spin hydrodynamics
has not been developed yet in our community. It is worthwhile to search
for analytic solutions for spin hydrodynamics in some simple, but
nevertheless realistic, test cases.

In this work, we study analytically the canonical relativistic spin
hydrodynamics in the time-honored longitudinal boost invariant Bjorken
flow \citep{Bjorken:1982qr}. We also consider the viscous effects
coupled to the spin hydrodynamics. Since relativistic hydrodynamics
in the first order of the gradient expansion are unstable and violate
the causality \citep{Hiscock:1983zz,Hiscock:1985zz,Hiscock:1987zz,Denicol:2008ha,Koide:2009sy,Pu:2009fj,Pu:2011vr},
we solve the equations of spin hydrodynamic coupled to the simplest
relaxation time corrections for the spin dynamics. To see the spin
corrections to the ordinary terms, we take the pseudo gauge transformation
and check the results in the Belinfante form of spin dynamics.

We emphasize that in the current work we are searching for the self-consistent
analytic solutions of dissipative spin hydrodynamics, which are different
with the studies of spin hydrodynamics in Bjorken \citep{Florkowski:2019qdp,Singh:2020efc,Singh:2021man}
and Gubser \citep{Singh:2020rht} expanding backgrounds.

The structure of this work is as following. In Sec. \ref{sec:Main-equations},
we will briefly review the conservation equations of dissipative spin
hydrodynamics in both canonical and Belinfante form. In Sec. \ref{subsec:Canonical-EMT-Bjorken},
we study the canonical spin hydrodynamics with non-vanishing viscous
tensor in a longitudinally boost invariant fluid and obtain the analytic
solutions. Next, we consider the relaxation time equations for spin
dynamics and derive the analytic solutions in the gradient expansion
in Sec. \ref{subsec:Relaxation-time}. We also take the pseudo gauge
transformation and compute the Belinfante form spin hydrodynamics
in Sec. \ref{subsec:Belinfante_EMT} and discuss our results in Sec.
\ref{subsec:Discussion}. We summarize in Sec. \ref{sec:Conclusion-and-discussion}.

Throughout this work, we adopt the metric $g^{\mu\nu}=\textrm{diag}\{+,-,-,-\}$
and the fluid velocity $u^{\mu}$ satisfying $u^{2}=1$. We define
the projector $\Delta^{\mu\nu}=g^{\mu\nu}-u^{\mu}u^{\nu}$. For simplicity,
for an arbitrary tensor $A^{\mu\nu}$, we introduce the symmetric,
anti-symmetric and traceless parts as $A^{(\mu\nu)}=(A^{\mu\nu}+A^{\nu\mu})/2$,
$A^{[\mu\nu]}=(A^{\mu\nu}-A^{\nu\mu})/2$ and $A^{\left\langle \mu\nu\right\rangle }=\frac{1}{2}[\Delta^{\mu\alpha}\Delta^{\nu\beta}+\Delta^{\nu\alpha}\Delta^{\mu\beta}]A_{\alpha\beta}-\frac{1}{3}\Delta^{\mu\nu}(A^{\rho\sigma}\Delta_{\rho\sigma})$.

\section{Main equations in relativistic dissipative spin hydrodynamics \label{sec:Main-equations}}

In this section, we briefly review the main equations of dissipative
spin hydrodynamics in both canonical and Belinfante form. In canonical
form of relativistic spin hydrodynamics, the conserved total angular
momentum (TAM) can be decomposed as \citep{Fukushima:2020ucl}
\begin{eqnarray}
J^{\alpha\mu\nu} & = & x^{\mu}T^{\alpha\nu}-x^{\nu}T^{\alpha\mu}+\Sigma^{\alpha\mu\nu},
\end{eqnarray}
where $T^{\mu\nu}$ is the the canonical energy momentum tensor (EMT)
and $\Sigma^{\alpha\mu\nu}$ is a rank-3 spin tensor corresponding
to the classical spin part of TAM. Up to $\mathcal{O}(\partial^{1})$
in the gradient expansion, the canonical EMT can be further decomposed
as \citep{Hattori:2019lfp,Fukushima:2020qta,Li:2020eon}, 
\begin{eqnarray}
T^{\mu\nu} & = & eu^{\mu}u^{\nu}-p\Delta^{\mu\nu}+2h^{(\mu}u^{\nu)}+2q^{[\mu}u^{\nu]}+\pi^{\mu\nu}+\phi^{\mu\nu},\label{eq:EMT_can}
\end{eqnarray}
where $e,p,u^{\mu},h^{\mu}$ and $\pi^{\mu\nu}$ denote energy density,
pressure, fluid velocity, heat flow and viscous tensor, respectively.
The $q^{\mu}$ and $\phi^{\mu\nu}$ are the anti-symmetric EMT parts.
Note that, $h^{\mu},q^{\mu},\pi^{\mu\nu}$ and $\phi^{\mu\nu}$ are
perpendicular to the fluid velocity $u^{\mu}$.

Both EMT and TAM are conserved, i.e. 
\begin{eqnarray}
\partial_{\mu}T^{\mu\nu} & = & 0,\nonumber \\
\partial_{\alpha}J^{\alpha\mu\nu} & = & 0,\label{eq:main_consev_01}
\end{eqnarray}
which lead to, 
\begin{eqnarray}
\partial_{\alpha}\Sigma^{\alpha\mu\nu} & = & -2T^{[\mu\nu]}.\label{eq:spin_tensor_diff}
\end{eqnarray}
It means the antisymmetric part of EMT acts as a source or absorption
term for spin current. Then we make a tensor decomposition for $\Sigma^{\alpha\mu\nu}$,
\begin{equation}
\Sigma^{\alpha\mu\nu}=u^{\alpha}S^{\mu\nu}+\Sigma_{(1)}^{\alpha\mu\nu},\label{eq:spin_tensor_01}
\end{equation}
where $S^{\mu\nu}$ is named as spin density and $\Sigma_{(1)}^{\alpha\mu\nu}$
refers to the higher order correction with $u_{\alpha}\Sigma_{(1)}^{\alpha\mu\nu}=0$.
Inserting Eqs. (\ref{eq:EMT_can}, \ref{eq:spin_tensor_01}) into
Eq. (\ref{eq:spin_tensor_diff}) yields, 
\begin{eqnarray}
\partial_{\alpha}(u^{\alpha}S^{\mu\nu}) & = & -2(q^{\mu}u^{\nu}-q^{\nu}u^{\mu}+\phi^{\mu\nu}).\label{eq:spin_tensor_02}
\end{eqnarray}

After introducing the spin degree of freedom , it is necessary to
modify the thermodynamic relations. We treat spin density $S^{\mu\nu}$
as particle number density, then we also need to introduce the spin
potential $\omega_{\mu\nu}=-\omega_{\nu\mu}$. Now, the new thermodynamic
relations become,
\begin{eqnarray}
e+p & = & Ts+\omega_{\mu\nu}S^{\mu\nu},\nonumber \\
de & = & Tds+\omega_{\mu\nu}dS^{\mu\nu},\nonumber \\
dp & = & sdT+S^{\mu\nu}d\omega_{\mu\nu},\label{eq:thermodynamic_01}
\end{eqnarray}
where $s$ is the entropy density and $T$ is the temperature. In
general, we can also have the charge currents in a system. For simplicity,
we neglect these currents throughout this work. For more general discussions,
one can also see Ref. \citep{Hattori:2019lfp,Fukushima:2020qta,Li:2020eon}.

The second law of thermodynamics requires that the entropy production
rate should always be positive definite. It gives the general expression
for those dissipative terms. The $q^{\mu}$ and $\phi^{\mu\nu}$ are
given by, 
\begin{eqnarray}
q^{\mu} & = & \lambda\left[\frac{1}{T}\Delta^{\mu\alpha}\partial_{\alpha}T+(u\cdot\partial)u^{\mu}-4\omega^{\mu\nu}u_{\nu}\right],\nonumber \\
\phi^{\mu\nu} & = & -\gamma\left(\Omega^{\mu\nu}-2\frac{1}{T}\Delta^{\mu\alpha}\Delta^{\nu\beta}\omega_{\alpha\beta}\right),\label{eq:phi_01}
\end{eqnarray}
where $\Omega^{\mu\nu}=-\Delta^{\mu\rho}\Delta^{\nu\sigma}\frac{1}{2}[\partial_{\rho}(\frac{u_{\sigma}}{T})-\partial_{\sigma}(\frac{u_{\rho}}{T})]$
is the thermal vorticity \citep{Becattini:2013fla,Fang:2016uds} and
$\lambda,\gamma\geq0$ are the transport coefficients. The ordinary
viscous tensor $\pi^{\mu\nu}$ is given by, 
\begin{eqnarray}
\pi^{\mu\nu} & = & \eta_{s}\partial^{\left\langle \mu\right.}u^{\left.\nu\right\rangle }-\Pi\Delta^{\mu\nu},\nonumber \\
\Pi & = & -\zeta(\partial\cdot u),
\end{eqnarray}
where $\eta_{s}$ and $\zeta$ are the shear and bulk viscosities,
respectively.

Note that, all of above results based on the assumption that the $\omega^{\mu\nu}$
and/or $S^{\mu\nu}$ are at $\mathcal{O}(\partial^{1})$. If both
$\omega^{\mu\nu}$ and $S^{\mu\nu}$ are at the leading order of the
gradient expansion, then the decomposition of EMT can be very different,
also see Ref. \citep{She:2021lhe}.

The canonical EMT is gauge dependent at the operator level and not
symmetric. One can choose Belinfante form EMT $\mathcal{T}^{\mu\nu}$,
which is gauge invariant and symmetric, through the following pseudo-gauge
transformation, 
\begin{eqnarray}
\mathcal{T}^{\mu\nu} & = & T^{\mu\nu}+\partial_{\lambda}K^{\lambda\mu\nu},
\end{eqnarray}
with
\begin{equation}
K^{\lambda\mu\nu}=\frac{1}{2}\left(\Sigma^{\lambda\mu\nu}-\Sigma^{\mu\lambda\nu}+\Sigma^{\nu\mu\lambda}\right).\label{eq:pesudo_gauge_01}
\end{equation}
According to the Ref. \citep{Fukushima:2020ucl}, up to $\mathcal{O}(\partial^{1})$,
we get,
\begin{eqnarray}
\mathcal{T}^{\mu\nu} & = & eu^{\mu}u^{\nu}-p\Delta^{\mu\nu}+\frac{1}{2}\partial_{\lambda}(u^{\mu}S^{\nu\lambda}+u^{\nu}S^{\mu\lambda})\nonumber \\
 & = & (e+\delta e)u^{\mu}u^{\nu}-p\Delta^{\mu\nu}+2(h^{(\mu}+\delta h^{(\mu})u^{\nu)}+(\pi^{\mu\nu}+\delta\pi^{\mu\nu}),
\end{eqnarray}
Here we have introduced the spin corrections to the energy density,
$\delta e$, the heat flow, $\delta h^{\mu}$ and the viscous tensor,
$\delta\pi^{\mu\nu}$, 
\begin{eqnarray}
\delta e & = & u_{\mu}\partial_{\lambda}S^{\mu\lambda},\nonumber \\
\delta h^{\mu} & = & \frac{1}{2}(\Delta_{\beta}^{\mu}\partial_{\lambda}S^{\beta\lambda}+u_{\beta}S^{\beta\lambda}\partial_{\lambda}u^{\mu}),\nonumber \\
\delta\pi^{\mu\nu} & = & \partial_{\lambda}(u^{\left\langle \mu\right.}S^{\left.\nu\right\rangle \lambda})+\frac{1}{3}\Delta_{\rho\sigma}\partial_{\lambda}(u^{\rho}S^{\sigma\lambda})\Delta^{\mu\nu}.\label{eq:delta_spin_01}
\end{eqnarray}
The Belinfante TAM becomes 
\begin{equation}
\mathcal{J}^{\alpha\mu\nu}=J^{\alpha\mu\nu}+\partial_{\rho}(x^{\mu}K^{\rho\alpha\nu}-x^{\nu}K^{\rho\alpha\mu})=x^{\mu}\mathcal{T}^{\alpha\nu}-x^{\nu}\mathcal{T}^{\alpha\mu}.
\end{equation}
Both Belinfante TAM and EMT are conserved. Also see Ref. \citep{Fukushima:2020ucl,Fukushima:2020qta}
for the discussion about the connection between canonical and Belinfante
form spin hydrodynamics.

\section{Analytic solutions in Bjorken expansion \label{sec:Analytic-solutions}}

In this section, we consider the time-honored Bjorken expansion of
ordinary relativistic hydrodynamics to the dissipative spin hydrodynamics.
The basic idea of Bjorken expansion is as follows. The fluid velocity
is given by \citep{Bjorken:1982qr}, 
\begin{equation}
u^{\mu}=\left(\frac{t}{\tau},0,0,\frac{z}{\tau}\right),\label{eq:Bjorken_01}
\end{equation}
where $\tau=\sqrt{t^{2}-z^{2}}$ is the proper time. The system is
assumed to be homogenous in transverse plane. As a consequence, all
of the macroscopic quantities only depend on the proper time $\tau$
and are independent on the space rapidity $\eta=\frac{1}{2}\ln[(t+z)/(t-z)]$.

To close the system, the equations of state (EoS) are essential. We
choose the energy density $e$ and $\omega^{\mu\nu}$ as two kinds
of thermodynamic variables. In this work, we follow the simplest EoS
for the relativistic fluid in the high temperature limit, 
\begin{equation}
e=3p.\label{eq:EoS_01}
\end{equation}
In Analogy to number density and chemical potential in the high temperature
limit, we assume the EoS for $S^{\mu\nu}$ is,
\begin{equation}
S^{\mu\nu}=a_{1}T^{2}\omega^{\mu\nu},\label{eq:EoS_omega_01}
\end{equation}
where $a_{1}$ is constant.

Similar to our previous works on searching for the analytic solutions
of relativistic magnetohydrodynamics \citep{Pu:2016ayh,Roy:2015kma,Pu:2016bxy,Pu:2016rdq,Siddique:2019gqh,Wang:2020qpx},
our strategy is as follows. As initial conditions, we assume that
the inital fluid velocity is given by Eq. (\ref{eq:Bjorken_01}) and
all the thermodynamic quantities are independent on $x,y$ and $\eta$.
Next, we will compute the dynamical evolution equations for fluid
velocity $u^{\mu}$, energy density $e$ and spin density $S^{\mu\nu}$.
At last, we search for the special configurations of $e$ and $\omega^{\mu\nu}$,
which can hold the initial Bjorken velocity (\ref{eq:Bjorken_01}).
To keep the whole system boost invariant, we therefore assume that
only $\omega^{xy}$ is nonzero initially. Then EoS (\ref{eq:EoS_omega_01})
reduces to,
\begin{equation}
S^{xy}=a_{1}T^{2}\omega^{xy}.\label{eq:EoS_omega_02}
\end{equation}

Furthermore, to simplify the calculations and to highlight the spin
effect, we choose the Landau frame in which heat flow $h^{\mu}$ is
always zero.

\subsection{Canonical form spin hydrodynamics in Bjorken flow \label{subsec:Canonical-EMT-Bjorken}}

Now, we discuss the canonical EMT and the corresponding conservation
equations for the Bjorken flow. The main equations are Eqs. (\ref{eq:main_consev_01},
\ref{eq:spin_tensor_02}).

Firstly, let us consider the the acceleration equation $\Delta_{\nu\alpha}\partial_{\mu}T^{\mu\nu}=0$,
which provides the dynamical evolution of fluid velocity,
\begin{eqnarray}
(u\cdot\partial)u_{\alpha} & = & \frac{1}{(e+p)}\left[\Delta_{\alpha}^{\mu}\partial_{\mu}p-(q\cdot\partial)u_{\alpha}+q_{\alpha}(\partial\cdot u)+\Delta_{\nu\alpha}(u\cdot\partial)q^{\nu}\right.\nonumber \\
 &  & \left.-\Delta_{\nu\alpha}\partial_{\mu}\phi^{\mu\nu}-\Delta_{\nu\alpha}\partial_{\mu}\pi^{\mu\nu}\right].\label{eq:du_01}
\end{eqnarray}
If $\omega^{xy}$ depends on the proper time $\tau$ only, with the
Bjorken velocity (\ref{eq:Bjorken_01}), the $q^{\mu}$ and $\phi^{\mu\nu}$
becomes,
\begin{eqnarray}
q^{\mu} & = & -4\lambda\omega^{\mu\nu}u_{\nu},\nonumber \\
\phi^{\mu\nu} & = & \frac{2\gamma}{T}[\omega^{\mu\nu}+2u^{[\mu}\omega^{\nu]\beta}u_{\beta}].\label{eq:q_phi_01}
\end{eqnarray}
Inserting Eq. (\ref{eq:q_phi_01}) into Eq. (\ref{eq:du_01}), we
find that the Bjorken velocity (\ref{eq:Bjorken_01}) will not be
modified under our assumption.

Next, let us consider the energy conservation equation $u_{\nu}\partial_{\mu}T^{\mu\nu}=0$,
i.e,
\begin{equation}
(u\cdot\partial)e+(e+p)\partial\cdot u+\partial\cdot q+q^{\nu}(u\cdot\partial)u_{\nu}+u_{\nu}\partial_{\mu}\phi^{\mu\nu}-\pi^{\mu\nu}\partial_{\mu}u_{\nu}=0.
\end{equation}
Using Eqs. (\ref{eq:q_phi_01}), the above equation becomes,
\begin{equation}
\frac{d}{d\tau}e+\frac{4}{3}e\frac{1}{\tau}-s\left(\frac{2}{3}\frac{\eta_{s}}{s}+\frac{\zeta}{s}\right)\frac{1}{\tau^{2}}=0,\label{eq:de_01}
\end{equation}
where the $s$ is the entropy. Note that, we assume that the dimensionless
quantities $\eta_{s}/s$ and $\zeta/s$ are constant.

Thirdly, from Eq. (\ref{eq:q_phi_01}), the spin density in Eq. (\ref{eq:spin_tensor_02})
reads,
\begin{eqnarray}
\frac{dS^{xy}}{d\tau}+S^{xy}\frac{1}{\tau} & = & -\frac{4\gamma}{T}\omega^{xy}.\label{eq:S_xy_01}
\end{eqnarray}
Note that, in this case, the $q^{\mu}$ does not contribute to the
$S^{xy}$.

To solve Eqs. (\ref{eq:de_01}, \ref{eq:S_xy_01}), we can express
the $s$ and $e$ as functions of $T$ and $\omega^{xy}$ by using
EoS (\ref{eq:EoS_01}, \ref{eq:EoS_omega_02}) and the thermodynamic
relations (\ref{eq:thermodynamic_01}), 
\begin{eqnarray}
e(T,\omega^{xy}) & = & c_{1}T^{4}+3a_{1}T^{2}\omega_{xy}^{2},\nonumber \\
s(T,\omega^{xy}) & = & \frac{4}{3}c_{1}T^{3}+2a_{1}T\omega_{xy}^{2},\label{eq:e_s_01}
\end{eqnarray}
where the constant $c_{1}$ can be determined by the the initial conditions,
i.e. 
\begin{equation}
c_{1}=\left[\frac{e_{0}}{T_{0}^{4}}-3a_{1}\left(\frac{\omega_{0}^{xy}}{T_{0}}\right)^{2}\right],
\end{equation}
with $e_{0}=e(\tau_{0})$, $T_{0}=T(\tau_{0})$ and $\omega_{0}^{xy}$
being the energy density, temperature and spin chemical potential
at the initial proper time $\tau_{0}$, respectively.

After inserting Eqs.(\ref{eq:e_s_01}) into Eq. (\ref{eq:de_01}),
we find that the solutions cannot be written into a compact form.
On the other hand, as mentioned in Sec. \ref{sec:Main-equations},
we have always assume that the $\omega^{xy}\sim\mathcal{O}(\partial^{1})$
in the gradient expansion in current formalism and it implies that
\begin{equation}
\omega^{xy}/T\ll1.
\end{equation}
Therefore, we can consider a power expansion of $\omega^{xy}/T$,
which is equivalent to the gradient expansion. Eqs. (\ref{eq:de_01},
\ref{eq:S_xy_01}) with Eqs. (\ref{eq:e_s_01}) becomes, 
\begin{eqnarray}
\frac{d}{d\tau}T+\frac{1}{3}\frac{T}{\tau}-\frac{1}{3}\left(\frac{2}{3}\frac{\eta_{s}}{s}+\frac{\zeta}{s}\right)\frac{1}{\tau^{2}}+\mathcal{O}\left((\omega_{xy}/T)^{2}\right) & = & 0,\nonumber \\
T\frac{d}{d\tau}\omega^{xy}+2\omega^{xy}\frac{d}{d\tau}T+T\omega^{xy}\frac{1}{\tau}+\frac{4\gamma}{a_{1}T^{2}}\omega^{xy}+\mathcal{O}\left((\omega_{xy}/T)^{2}\right) & = & 0.
\end{eqnarray}
The solutions are, 
\begin{eqnarray}
T(\tau) & = & T_{0}\left(\frac{\tau_{0}}{\tau}\right)^{1/3}-\frac{1}{2\tau}\left(\frac{2}{3}\frac{\eta_{s}}{s}+\frac{\zeta}{s}\right)\left[1-\left(\frac{\tau}{\tau_{0}}\right)^{2/3}\right]+\mathcal{O}\left((\omega_{0}^{xy}/T_{0})^{2}\right),\nonumber \\
\omega^{xy}(\tau) & = & \omega_{0}^{xy}\left(\frac{\tau_{0}}{\tau}\right)^{1/3}\exp\left[-\frac{2\gamma\tau_{0}}{a_{1}T_{0}^{3}}\left(\frac{\tau^{2}}{\tau_{0}^{2}}-1\right)\right]\left\{ 1+\left(\frac{2}{3}\frac{\eta_{s}}{s}+\frac{\zeta}{s}\right)\frac{1}{T_{0}^{4}}\right.\nonumber \\
 &  & \left.\times\left[\frac{T_{0}^{3}}{\tau_{0}}\left(\left(\frac{\tau_{0}}{\tau}\right)^{2/3}-1\right)+\frac{\gamma}{a_{1}}\left(3\left(\frac{\tau}{\tau_{0}}\right)^{2}-\frac{9}{2}\left(\frac{\tau}{\tau_{0}}\right)^{4/3}+\frac{3}{2}\right)\right]\right\} \nonumber \\
 &  & +\mathcal{O}\left((\omega_{0}^{xy}/T_{0})^{2},(\eta_{s}/s)^{2},(\zeta/s)^{2},(\eta_{s}\zeta/s^{2})\right),\label{eq:sol_01}
\end{eqnarray}
where for $\omega^{xy}(\tau)$, we only keep the linear terms of $\eta_{s}/s$
and $\zeta/s$.

Inserting solutions (\ref{eq:sol_01}) into Eqs. (\ref{eq:e_s_01})
and EoS (\ref{eq:EoS_omega_01}) yields, 
\begin{eqnarray}
e(\tau) & = & e_{0}\left(\frac{\tau_{0}}{\tau}\right)^{4/3}-2\frac{e_{0}\tau_{0}}{T_{0}\tau^{2}}\left(\frac{2}{3}\frac{\eta_{s}}{s}+\frac{\zeta}{s}\right)\left[1-\left(\frac{\tau}{\tau_{0}}\right)^{2/3}\right]\nonumber \\
 &  & +\mathcal{O}\left((\omega_{0}^{xy}/T_{0})^{2},(\eta_{s}/s)^{2},(\zeta/s)^{2},(\eta_{s}\zeta/s^{2})\right),
\end{eqnarray}
and 
\begin{eqnarray}
S^{xy}(\tau) & = & a_{1}\omega_{0}^{xy}T_{0}^{2}\left(\frac{\tau_{0}}{\tau}\right)\exp\left[-\frac{2\gamma\tau_{0}}{a_{1}T_{0}^{3}}(\frac{\tau^{2}}{\tau_{0}^{2}}-1)\right]\nonumber \\
 &  & \times\left\{ 1+\left(\frac{2}{3}\frac{\eta_{s}}{s}+\frac{\zeta}{s}\right)\frac{3\gamma}{2a_{1}T_{0}^{4}}\left[2\left(\frac{\tau}{\tau_{0}}\right)^{2}-3\left(\frac{\tau}{\tau_{0}}\right)^{\frac{4}{3}}+1\right]\right\} \nonumber \\
 &  & +\mathcal{O}\left((\omega_{0}^{xy}/T_{0})^{2},(\eta_{s}/s)^{2},(\zeta/s)^{2},(\eta_{s}\zeta/s^{2})\right).\label{eq:sol_S_01}
\end{eqnarray}

The dissipative effects from $\phi^{\mu\nu}$ contribute an extra
factor $\exp[-2\gamma(\tau^{2}-\tau_{0}^{2})/(a_{1}T_{0}^{3}\tau_{0})]$
to Eqs. (\ref{eq:sol_01}, \ref{eq:sol_S_01}). Both $\omega^{xy}$
and $S^{xy}$ decay much rapidity in a finite $\phi^{\mu\nu}$ system
than in a vanishing $\phi^{\mu\nu}$ system. Note that the factor
related to $\gamma$ looks similar to the factor $\sim\exp[-\sigma(\tau-\tau_{0})]$
caused by the electric conducting flow in magnetohydrodynamics with
$\sigma$ the electric conductivity \citep{Siddique:2019gqh}. From
Eq. (\ref{eq:sol_S_01}), we find that the viscous corrections to
$S^{xy}(\tau)$ are always positive when $\tau/\tau_{0}\geq1$ and
increase when $\tau_{0}$ grows. It means that the viscous effects
accelerate the decay of spin density.

\subsection{Corrections from relaxation time for spin transport \label{subsec:Relaxation-time}}

In this section, we compute the higher order effects for the spin
transport. For simplicity, we neglect $\pi^{\mu\nu}$ and concentrate
on the spin effects. It is well-known that the relativistic dissipative
hydrodynamics in $\mathcal{O}(\partial^{1})$ violate causality and
are unstable, e.g. see Ref. \citep{Hiscock:1983zz,Hiscock:1985zz,Hiscock:1987zz,Denicol:2008ha,Koide:2009sy,Pu:2009fj,Pu:2011vr}
and the references therein. Therefore, one needs to introduce the
second order corrections. In this work, we only consider a standard
second order correction to the dissipative terms in Eqs. (\ref{eq:phi_01}),
\begin{equation}
\tau_{\phi}\frac{d}{d\tau}\phi^{\mu\nu}+\phi^{\mu\nu}=-\gamma\left(\Omega^{\mu\nu}-2\frac{1}{T}\Delta^{\mu\alpha}\Delta^{\nu\beta}\omega_{\alpha\beta}\right),\label{eq:RTA_01}
\end{equation}
where $\tau_{\phi}$ is the relaxation time for $\phi^{\mu\nu}$.
The relaxation time $\tau_{\phi}$ describes how fast the system could
reach to the equilibrium again after taking some perturbations $\phi^{\mu\nu}$
to an equilibrium system. It is similar to the relaxation time equations
for viscous terms. Also see Ref. \citep{Hattori:2020htm,Gallegos:2021bzp,Li:2020eon,She:2021lhe}
for other possible second order corrections of spin hydrodynamics.

In this case, the conservation equations become, 
\begin{eqnarray}
\frac{d}{d\tau}e+\frac{4}{3}e\frac{1}{\tau} & = & 0,\label{eq:e_02}
\end{eqnarray}
and, 
\begin{equation}
(u\cdot\partial)u_{\alpha}=\frac{1}{(e+p)}[\Delta_{\alpha}^{\mu}\partial_{\mu}p-\Delta_{\nu\alpha}\partial_{\mu}\phi^{\mu\nu}].\label{eq:acc_u_02}
\end{equation}
The evolution equations for the spin density become, 
\begin{eqnarray}
\frac{dS^{xy}}{d\tau}+S^{xy}\frac{1}{\tau} & = & -2\phi^{xy},\nonumber \\
\tau_{\phi}\frac{d}{d\tau}\phi^{xy}+\phi^{xy} & = & \frac{2\gamma}{T}\omega^{xy}.\label{eq:S_02}
\end{eqnarray}
We find that the fluid velocity will not be modified when $\phi^{xy}$
only depends on the $\tau$. It is straightforward to get the solution
for Eq. (\ref{eq:e_02}), 
\begin{eqnarray}
e(\tau) & = & e_{0}\left(\frac{\tau_{0}}{\tau}\right)^{4/3},\nonumber \\
T(\tau) & = & T_{0}\left(\frac{\tau_{0}}{\tau}\right)^{1/3}+\mathcal{O}\left((\omega_{0}^{xy}/T_{0})^{2}\right).
\end{eqnarray}
Then, we can solve the $\phi^{xy},S^{xy}$ and $\omega^{xy}$, 
\begin{eqnarray}
\phi^{xy}(\tau) & = & e^{-\frac{\tau}{2\tau_{\phi}}}f(\tau)+\mathcal{O}\left((\omega_{0}^{xy}/T_{0})^{2}\right),\nonumber \\
\omega^{xy}(\tau) & = & \left(\frac{\tau_{0}}{\tau}\right)^{\frac{1}{3}}\frac{T_{0}}{2\gamma}e^{-\frac{\tau}{2\tau_{\phi}}}\left[\frac{1}{2}f(\tau)-\left(\frac{4\gamma\tau_{\phi}^{2}}{a_{1}\tau_{0}T_{0}^{3}}\right)^{\frac{1}{3}}g(\tau)\right]+\mathcal{O}\left((\omega_{0}^{xy}/T_{0})^{2}\right),\nonumber \\
S^{xy}(\tau) & = & a_{1}\left(\frac{\tau_{0}}{\tau}\right)\frac{T_{0}^{3}}{2\gamma}e^{-\frac{\tau}{2\tau_{\phi}}}\left[\frac{1}{2}f(\tau)-\left(\frac{4\gamma\tau_{\phi}^{2}}{a_{1}\tau_{0}T_{0}^{3}}\right)^{\frac{1}{3}}g(\tau)\right]+\mathcal{O}\left((\omega_{0}^{xy}/T_{0})^{2}\right),\label{eq:omega_02}
\end{eqnarray}
where $f(\tau)$ and $g(\tau)$ are shown in Appendix \ref{sec:Expression}.

In the $\tau_{\phi}\rightarrow0$ limit, we prove that the solutions
(\ref{eq:omega_02}) reduce to Eqs. (\ref{eq:sol_01}, \ref{eq:sol_S_01}).
In the $\tau_{\phi}\rightarrow\infty$ limit, the solutions (\ref{eq:omega_02})
become, 
\begin{equation}
\lim_{\tau_{\phi}\rightarrow\infty}\omega^{xy}(\tau)=\omega_{0}^{xy}(\tau_{0}/\tau)^{1/3},\qquad\lim_{\tau_{\phi}\rightarrow\infty}S^{xy}(\tau)=a_{1}T_{0}^{2}\omega_{0}^{xy}(\tau_{0}/\tau),.
\end{equation}
Therefore, the new term proportional to $\tau_{\phi}$ in Eq. (\ref{eq:RTA_01})
slows down the decay caused by the $\phi^{\mu\nu}$ as expected.

\subsection{Results for Belinfante form EMT \label{subsec:Belinfante_EMT}}

By using the pseudo gauge transformation (\ref{eq:pesudo_gauge_01}),
we can obtain the Belinfante form EMT $\mathcal{T}^{\mu\nu}$. With
the solutions (\ref{eq:sol_S_01}) in Sec. \ref{subsec:Canonical-EMT-Bjorken}
and solutions (\ref{eq:omega_02}) in Sec. \ref{subsec:Relaxation-time},
we find that 
\begin{eqnarray}
\frac{1}{2}\partial_{\lambda}(u^{\mu}S^{\nu\lambda}+u^{\nu}S^{\mu\lambda}) & = & \mathcal{O}(\partial^{2}),
\end{eqnarray}
up to the order of $\mathcal{O}(\partial^{2})$. Therefore, Belinfante
EMT reduces to the ordinary EMT without spin effects. We have checked
these spin corrections in Eqs. (\ref{eq:delta_spin_01}) and found
that all of them vanish under our assumption.

We emphasize that it does not mean that there are no the spin effects
in Belinfante form of dissipative spin hydrodynamics. In order to
observe these spin corrections in Eqs. (\ref{eq:delta_spin_01}),
we have to consider the inhomogeneity in the transverse plane.

\subsection{Discussion \label{subsec:Discussion}}

Let us take a close look to the leading order of $\omega^{xy}$ and
$S^{xy}$ in Eqs. (\ref{eq:sol_01}, \ref{eq:sol_S_01}). We find
that $\omega^{xy}$ and $S^{xy}$ decay as $\sim\tau^{-1/3}$ and
$\sim1/\tau$, respectively. From Ref. \citep{Becattini:2013fla,Fang:2016uds},
one can compute the spin polarization of $\Lambda$ hyperons caused
by the thermal vorticity by the modified Cooper-Frye formula, 
\begin{equation}
\mathcal{S}^{\mu}({\bf p})=\frac{\int d\Sigma\cdot pf(1-f)\epsilon^{\mu\nu\alpha\beta}p_{\nu}\left[\partial_{\alpha}\left(\frac{u_{\beta}}{T}\right)+...\right]}{8m_{\Lambda}\int d\Sigma\cdot pf},
\end{equation}
where the $m_{\Lambda}$ is the mass of $\Lambda$ hyperon, $\Sigma^{\mu}$
is the freezeout hypersurface, $f=f(x,p)$ is the distribution functions
of particles and $...$ denotes the other possible corrections from
shear viscous tensor and others \citep{Hidaka:2018mel,Liu:2020dxg,Liu:2021uhn,Fu:2021pok,Becattini:2021iol,Yi:2021ryh}.
In general, the spin density $S^{\mu\nu}$ and $\omega^{\mu\nu}$
contribute to the distribution function $f(x,p)$. Our results (\ref{eq:sol_01},
\ref{eq:sol_S_01}) imply that the initial spin density decays rapidly
and may not play an important role to the polarization vector $\mathcal{S}^{\mu}({\bf p})$
at the freezeout hypersurface. Although the huge initial orbital angular
momentum could transfer to the spin, it may still be difficult to
accumulate the net spin density in such an expanding system. Therefore,
one may still neglect the contributions from $S^{\mu\nu}$ or $\omega^{\mu\nu}$
to the calculations of polarization vector $\mathcal{S}^{\mu}({\bf p})$.

\section{Conclusion \label{sec:Conclusion-and-discussion}}

In this work, we have derived the analytic solutions of a longitudinally
boost-invariant dissipative spinness fluid with finite viscous tensor
and second order relaxation time corrections for the spin. Our main
results are shown in Eqs. (\ref{eq:sol_01}, \ref{eq:sol_S_01}) and
(\ref{eq:omega_02}).

We find that the spin density and spin chemical potential decay as
$\sim\tau^{-1}$ and $\tau^{-1/3}$, respectively. It implies that
the initial spin density may not survive at the late time. Although
the huge initial orbital angular momentum could transfer to the spin,
it may be difficult to accumulate the net spin density in such an
expanding system.

We have also computed the Belinfante EMT and have not found any corrections
from spin in this work. To observe the possible spin corrections to
the ordinary terms in Eqs. (\ref{eq:delta_spin_01}), one needs to
consider the inhomogeneity in the transverse plane, e.g. similar to
the Ref. \citep{Pu:2016bxy} for the magnetohydrodynamics.
\begin{acknowledgments}
D.L. Wang and S. Fang are grateful to the School of Science in Huzhou
University for its hospitality during the completion of this work.
This work is supported by National Nature Science Foundation of China
(NSFC) under Grants No. 12075235.
\end{acknowledgments}

\appendix

\section{Expression for Eqs. (\ref{eq:omega_02}) \label{sec:Expression}}

The $f(\tau)$ and $g(\tau)$ in Eqs. (\ref{eq:omega_02}) are 
\begin{eqnarray}
f(\tau) & = & C_{1}\mathrm{Ai}\left[\mathcal{H}(\tau)\right]+C_{2}\mathrm{Bi}\left[\mathcal{H}(\tau)\right],\nonumber \\
g(\tau) & = & C_{1}\mathrm{Ai}^{\prime}\left[\mathcal{H}(\tau)\right]+C_{2}\mathrm{Bi}^{\prime}\left[\mathcal{H}(\tau)\right],
\end{eqnarray}
where 
\begin{equation}
\mathcal{H}(\tau)=2^{-\frac{4}{3}}\left(\frac{1}{4\tau_{\phi}^{2}}-\frac{4\gamma\tau}{a_{1}\tau_{0}\tau_{\phi}T_{0}^{3}}\right)\left(-\frac{\gamma}{a_{1}\tau_{0}T_{0}^{3}\tau_{\phi}}\right)^{-\frac{2}{3}},
\end{equation}
The $\mathrm{Ai}(z)$ and $\mathrm{Bi}(z)$ are named as Airy functions,
which are the solutions of the differential equation $y^{\prime\prime}(z)\pm zy(z)=0$,
respectively. Note that, here both $\mathrm{Ai}(z)$ and $\mathrm{Bi}(z)$
are entire Airy functions of $z$ with no branch cut discontinuities.
We also use the notation $\mathrm{Ai}^{\prime}(x)=\frac{d}{dx}\mathrm{Ai}(x)$
and $\mathrm{Bi}^{\prime}(x)=\frac{d}{dx}\mathrm{Bi}(x)$.

The coefficients $C_{1,2}$ are determined by the initial condition
$\phi^{xy}(\tau_{0})=\phi_{0}^{xy}$ and $\omega^{xy}(\tau_{0})=\omega_{0}^{xy}$,
\begin{eqnarray}
C_{1} & = & \mathcal{A}^{-1}e^{\frac{\tau_{0}}{2\tau_{\phi}}}\left[\mathrm{Bi}(\mathcal{U})\mathcal{C}-\phi_{0}^{xy}\mathrm{Bi^{\prime}}(\mathcal{U})\right],\nonumber \\
C_{2} & = & \mathcal{A}^{-1}e^{\frac{\tau_{0}}{2\tau_{\phi}}}\left[\mathrm{Ai}^{\prime}(\mathcal{U})\phi_{0}^{xy}-\mathrm{Ai}(\mathcal{U})\mathcal{C}\right],
\end{eqnarray}
where 
\begin{eqnarray}
\mathcal{A} & = & \mathrm{Ai^{\prime}}(\mathcal{U})\mathrm{Bi}(\mathcal{U})-\mathrm{Ai}(\mathcal{U})\mathrm{Bi^{\prime}}(\mathcal{U}),\nonumber \\
\mathcal{C} & = & \left(\frac{a_{1}\tau_{0}T_{0}^{3}}{4\gamma\tau_{\phi}^{2}}\right)^{\frac{1}{3}}\left(\frac{1}{2}\phi_{0}^{xy}-\frac{2\gamma}{T_{0}}\omega_{0}^{xy}\right),\nonumber \\
\mathcal{U} & = & \left(\frac{a_{1}\tau_{0}T_{0}^{3}}{4\gamma}\right)^{\frac{2}{3}}\left(\frac{1}{4}\tau_{\phi}^{-\frac{4}{3}}-\frac{4\gamma}{a_{1}T_{0}^{3}}\tau_{\phi}^{-\frac{1}{3}}\right).
\end{eqnarray}
Note that, here we only take the real solutions and neglect imaginary
solutions.

\bibliographystyle{h-physrev}
\bibliography{spinhydro}

\end{document}